\documentclass[%
 aip,
 amsmath,amssymb,
reprint,
twocolumn,
]{revtex4-2}
\usepackage[english]{babel}
\usepackage{hyperref}
\usepackage{graphicx}
\usepackage{dcolumn}
\usepackage{bm}

\usepackage[utf8]{inputenc}
\usepackage[T1]{fontenc}
\usepackage{mathptmx}
\usepackage{amssymb}
\usepackage{mathrsfs}
\usepackage{color}
\begin{document}

\title[]{Nonadiabatic ab initio molecular dynamics
including spin–orbit coupling and laser fields}

\author{Philipp Marquetand} 
\email{philipp.marquetand@univie.ac.at}
\affiliation{%
Institut f\"ur Physikalische Chemie, Friedrich-Schiller-Universit\"at Jena, Helmholtzweg 4, 07743
Jena, Germany}
\author{Martin Richter}
\affiliation{%
Institut f\"ur Physikalische Chemie, Friedrich-Schiller-Universit\"at Jena, Helmholtzweg 4, 07743
Jena, Germany}
\author{Jes\'us Gonz\'alez-V\'azquez}
\affiliation{%
Institut f\"ur Physikalische Chemie, Friedrich-Schiller-Universit\"at Jena, Helmholtzweg 4, 07743
Jena, Germany}
\altaffiliation{%
Departamento de Qu\'imica F\'isica I, Universidad Complutense, 28040 Madrid, Spain}
\author{Ignacio Sola}
\affiliation{%
Departamento de Qu\'imica F\'isica I, Universidad Complutense, 28040 Madrid, Spain}
\author{Leticia Gonz\'alez}
\affiliation{%
Institut f\"ur Physikalische Chemie, Friedrich-Schiller-Universit\"at Jena, Helmholtzweg 4, 07743
Jena, Germany}

\date{\today}

\begin{abstract}
Nonadiabatic ab initio molecular dynamics (MD) including spin-orbit coupling (SOC) and laser fields is investigated as a general tool for studies of excited-state processes. Up to now, SOCs are not included in standard ab initio MD packages. Therefore, transitions to triplet states cannot be treated in a straightforward way. Nevertheless, triplet states play an important role in a large variety of systems and can now be treated within the given framework. The laser interaction is treated on a non-perturbative level that allows nonlinear effects like strong Stark shifts to be considered. As MD allows for the handling of many atoms, the interplay between triplet and singlet states of large molecular systems will be accessible. In order to test the method, IBr is taken as a model system, where $S O C$ plays a crucial role for the shape of the potential curves and thus the dynamics. Moreover, the influence of the nonresonant dynamic Stark effect is considered. The latter is capable of controlling reaction barriers by electric fields in timereversible conditions, and thus a control laser using this effect acts like a photonic catalyst. In the IBr molecule, the branching ratio at an avoided crossing, which arises from $\mathrm{SOC}$, can be influenced.
\end{abstract}
\maketitle

\section{\label{sec:Introduction}Introduction}

Laser control of chemical reactions has been a target for researchers for decades. The will to exert control goes hand in hand with the desire to witness the corresponding molecular processes in real time. Since the seminal works by Zewail and coworkers,,$^{1,2}$ observation of atomic motion in the femtosecond regime is possible. Nonetheless, it is still challenging to unravel complex processes in small and big systems including several electronic states. A joint effort of experiment and theory is necessary to achieve this goal. On the one hand, the theoretical description with atomistic models particularly helps to understand the complex processes as the calculations are directly carried out and visualized in the descriptive picture that we hold in our heads. On the other hand, the use of exact formulas is only possible for the simplest systems. ${ }^{3}$ The solution of the time-dependent Schrödinger equation to represent the dynamics of molecules in full dimensionality is not feasible for most systems with today's computers. Therefore, different approximations have been established and methods like e.g. Multiconfigurational time-dependent Hartree method (MCTDH), ${ }^{4-6}$ multiple spawning ${ }^{7,8}$ or similar techniques ${ }^{9-17}$ emerged.

Here, we want to focus on another ansatz, ab initio molecular dynamics (MD), ${ }^{18,19}$ where Newton's classical equations of motion are used to approximate the change of the nuclear positions while the electronic structure of the considered system is treated quantum mechanically. The gap between these two descriptions for different parts of the system is bridged by the surface hopping $(\mathrm{SH})$ method. ${ }^{20,21}$ However, the latter was originally developed to account for nonadiabatic couplings in the photodynamics of molecules. Recent efforts try to describe other types of coupling which may occur during or after photoexcitations. ${ }^{22-26,58,59}$ In a recent paper, we introduced the surface-hopping-in-adiabatic-representation-including-arbitrarycouplings (SHARC) method. ${ }^{27}$ The advantage of the latter method is that all kinds orbit coupling (SOC), laser excitations and nonadiabatic coupling can be modeled for complex systems with an arbitrary number of atoms and a large number of degrees of freedom.

Especially the interaction between molecules and laser fields is important in quantum control. ${ }^{28-38}$ In this contribution, we shall focus on a special type of control via the nonresonant dynamic Stark effect (NRDSE). ${ }^{39-42}$ The Stark effect, which is ubiquitious in strong field control, see e.g. Ref. 43 , is used to create dressed states, also called light-induced potentials (LIPs). Remarkable about NRDSE is that the laser field only shifts the potential via interactions with the permanent dipole moment and polarizability without inducing any resonant transitions. Therefore, reaction barriers can be altered with the NRSDE acting like a photonic catalyst. $^{39-42}$

As a model system, we have chosen $\mathrm{IBr}$ since it exhibits strong $\mathrm{SOC}$ leading to an avoided crossing between two excited states, the $1^{3} \Pi_{0+}$ and the $1^{3} \Sigma_{0+}^{-}$state, see Ref. 44 and references therein. Excited state dissociation in these two states leads to different product channels resulting in $\mathrm{I}+\mathrm{Br}$ and $\mathrm{I}+\mathrm{Br}^{*}$, respectively. The asterisk indicates the ${ }^{2} P_{1 / 2}$ excited spin-state of the dissociating $\mathrm{Br}$ atom, while the ground state has the configuration ${ }^{2} P_{3 / 2}$. After electronic excitation of the IBr molecule with a resonant laser, the branching ratio in the two channels is influenced with a second, nonresonant laser. We show that this NRDSE process can be described within the SHARC formalism. Results shall be compared to exact quantum dynamics (QD) calculations.

The methodology and the theoretical description are presented in Sec. 2; the numerical results are contained in Sec. 3, and a summary is given in Sec. 4. 

\section{Methodology}
We use ab initio MD, where the electrons of a molecular system are treated quantum mechanically and the nuclei classically. On the one hand, the nuclei follow classical trajectories defined by the nuclear position $\vec{R}(\mathrm{t})$ and velocity $\vec{v}(t)$ at every time following the quantum potential created by the electrons. On the other hand, the quantum potential is evaluated as the expectation value of an effective Hamiltonian $V(t)=\left\langle\Psi(\vec{R}(t) ; \vec{r}, t)\left|\hat{H}_{\mathrm{eff}}[\vec{R}(t) ; \vec{r}]\right| \Psi(\vec{R}(t) ; \vec{r}, t)\right\rangle$ and thus, depends parametrically on the nuclear coordinates.

To account for the quantum position-momentum uncertainty in the classical part of our simulations, we create a swarm of trajectories with different initial conditions. The latter are prepared in a way that resembles the probability distribution of the corresponding quantum wavepacket. Every single trajectory is propagated using the Velocity Verlet algorithm ${ }^{45,46}$ for the solution of Newton's equations. In this algorithm, the dynamics of the nuclear coordinates $\vec{R}(t)$ is governed by the gradient of the potential at time $t$ :
$$
\vec{R}(t+\Delta t)=\vec{R}(t)+\vec{v}(t) \Delta t+\frac{1}{2 M} \nabla_{\vec{R}} V(t) \Delta t^{2}
$$
Here, $M$ represents the mass of the nuclei and $\vec{v}(t)$ is the velocity of the trajectory. The time evolution of this velocity is calculated using the gradient of the potential at times $t$ and $t+\Delta t$ :
$$
\vec{v}(t+\Delta t)=\vec{v}(t)+\frac{1}{2 M} \nabla_{\vec{R}} V(t) \Delta t+\frac{1}{2 M} \nabla_{\vec{R}} V(t+\Delta t) \Delta t
$$
The definition of the potential is given by the time evolution of the electronic wavepacket following the time-dependent Schrödinger equation:
$$
i \hbar \frac{\partial|\Psi[\vec{R}(t) ; \vec{r}, t]\rangle}{\partial t}=\hat{H}_{\mathrm{eff}}[\vec{R}(t) ; \vec{r}]|\Psi[\vec{R}(t) ; \vec{r}, t]\rangle .
$$
We employ a linear basis set expansion of the wavepacket to solve this equation with implicitly time-dependent basis functions at different $\vec{R}(t)$ :
$$
|\Psi[\vec{R}(t) ; \vec{r}, t]\rangle=\sum_{m} c_{m}(t)\left|\phi_{m}[\vec{R}(t) ; \vec{r}]\right\rangle
$$
where $c_{m}$ are the amplitudes of the eigenfunctions $\phi_{m}$.

Consequently, the time evolution of these amplitudes are given by:
$$
\frac{\partial c_{n}(t)}{\partial t}=-\sum_{m}\left\{\frac{i}{\hbar} \mathrm{H}_{n m}[\vec{R}(t)]+\mathrm{K}_{n m}[\vec{R}(t)]\right\} c_{m}(t)
$$
The first term $\mathrm{H}_{n m}[\vec{R}(t)]=\left\langle\phi_{n}[\vec{R}(t) ; \vec{r}]\left|\hat{H}_{\mathrm{eff}}[\vec{R}(t) ; \vec{r}]\right| \phi_{m}[\vec{R}(t) ; \vec{r}]\right\rangle$ of this equationsolved by a simple Runge-Kutta algorithm of fourth order-describes the diabatic Hamiltonian with the different potentials as diagonal elements and the diabatic couplings as the off-diagonal elements. The second term, $\mathrm{K}_{n m}[\vec{R}(t)]$, represents the change of the electronic basis functions with time, which is equivalent to the variation of the basis with the nuclear coordinates multiplied by the velocity:
$$
\begin{aligned}
&\mathrm{K}_{n m}[\vec{R}(t)]=\left\langle\phi_{n}[\vec{R}(t) ; \vec{r}]|\partial / \partial t| \phi_{m}[\vec{R}(t) ; \vec{r}]\right\rangle \\
&\quad=\left\langle\phi_{n}[\vec{R}(t) ; \vec{r}]|d / d \vec{R}(t)| \phi_{m}[\vec{R}(t) ; \vec{r}]\right\rangle \vec{v}(t)
\end{aligned}
$$
If the basis functions are chosen to be the eigenfunctions of the time-independent Schrödinger equation for every $\vec{R}(t)$, then the amplitudes are directly correlated with the populations of the different electronic states. This is important since a trajectory can only be influenced by a single potential at a time and the corresponding state has to be assigned. Here, we use Tully's SH method, ${ }^{20}$ which was originally developed to account for nonadiabatic couplings by giving the trajectory the possibility to jump from one state to another. The probability for such a hop is calculated using the time-dependent amplitudes introduced above:
$$
P_{n m}=\frac{2 \mathscr{R}\left\{c_{n}^{*}(t) c_{m}(t)\left[\frac{i}{\hbar} \mathrm{H}_{n m}[\vec{R}(t)]+\mathrm{K}_{n m}[\vec{R}(t)]\right]\right\}}{c_{n}^{*}(t) c_{n}(t)} \Delta t,
$$
where $\mathscr{R}$ denotes the real part.

The original SH methodology is widely used to simulate problems with localized couplings originating from conical intersections. ${ }^{47,48}$ In this work, we extend $\mathrm{SH}$ to the situation where different types of couplings, e.g. SOCs and/or the interaction with an electric field must be taken into account. These coupling terms are typically evaluated in the diabatic representation and hence included in the potential part of the Hamiltonian. In this way, a new $\mathrm{H}^{d}[\vec{R}(t)]$ matrix is introduced with elements (where the index $d$ indicates that additional nondiagonal terms are included):
$$
\mathrm{H}_{n m}^{d}[\vec{R}(t), t]=\mathrm{H}_{n m}[\vec{R}(t)]+\mathrm{H}_{n m}^{\mathrm{AC}}[\vec{R}(t)]
$$
In this equation, $\mathrm{H}_{n m}^{\mathrm{AC}}[\vec{R}(t), t]$ indicates the arbitrary-coupling matrix which can consist of terms like $-\vec{\mu}_{n m}[\vec{R}(t)] \vec{E}(t)+\mathrm{H}_{n m}^{S O}[\vec{R}(t)]$, where $\vec{\mu}_{n m}[\vec{R}(t)]$ and $\mathrm{H}_{n m}^{\mathrm{SO}}[\vec{R}(t)]$ are the dipole moment and the relativistic SOC between the states $n$ and $m$, respectively.

Such additional terms may complicate the solution of the corresponding equations since the interactions are mostly delocalized in contrast to the rather localized nonadiabatic couplings. In SHARC, we solve this problem by translating the additional elements to the $\mathrm{K}[\vec{R}(t)]$ matrix. We choose the adiabatic (index $a$ ) representation, where the $\mathrm{H}^{d}[\vec{R}(t), t]$ matrix is diagonalized and afterwards, the $\mathrm{K}[\vec{R}(t)]$ matrix, localizing the couplings in geometries where the electronic states are degenerated, is recalculated.

Along these lines, the basis set of electronic wavefunctions $\left|\phi^{d}[\vec{R}(t) ; r]\right\rangle$ is substituted for a linear combination,
$$
\left|\phi_{n}^{a}[\vec{R}(t) ; \vec{r}, t]\right\rangle=\sum_{m} \mathrm{U}_{n m}[\vec{R}(t), t]\left|\phi_{m}^{d}[\vec{R}(t) ; \vec{r}]\right\rangle,
$$
where $\mathrm{U}[\vec{R}(t), t]$ is the unitary matrix that diagonalizes the Hamiltonian $\mathrm{H}^{d}[\vec{R}(t), t]$ matrix at every time $t$. In this new basis, the elements of the $\mathrm{H}^{a}[\vec{R}(t), t]$ matrix are defined as:
$$
\mathrm{H}_{n m}^{a}[\vec{R}(t), t]=\mathrm{V}_{m}^{a}[\vec{R}(t), t] \delta_{n m}
$$
where $\mathrm{V}_{m}^{a}[\vec{R}(t), t]$ are the diagonal elements of $\mathrm{H}^{a}[\vec{R}(t), t]$. The couplings are treated via the derivative of the $\left|\phi^{a}[\vec{R}(t) ; \vec{r}, t]\right\rangle:$
$$
\begin{aligned}
\mathrm{K}_{n m}^{a}[\vec{R}(t), t] &=\left\langle\phi_{n}^{a *}[\vec{R}(t) ; \vec{r}, t]\left|\frac{\partial}{\partial t}\right| \phi_{m}^{a}[\vec{R}(t) ; \vec{r}, t]\right\rangle \\
&=\mathrm{K}_{n m}^{\phi}[\vec{R}(t), t]+\mathrm{K}_{n m}^{U}[\vec{R}(t), t]
\end{aligned}
$$
Application of the chain rule easily yields the definitions for these latter terms:
$$
\mathrm{K}_{n m}^{\phi}[\vec{R}(t), t]=\sum_{l k} \mathrm{U}_{l n}^{*}[\vec{R}(t), t] \mathrm{K}_{l k}[\vec{R}(t)] \mathrm{U}_{k m}[\vec{R}(t), t]
$$
which is just the rotation of the original nonadiabatic term to the new basis and
$$
\mathrm{K}_{n m}^{\mathrm{U}}[\vec{R}(t), t]=\sum_{l} \mathrm{U}_{l n}^{*}[\vec{R}(t), t] \frac{\partial}{\partial t} \mathrm{U}_{l m}[\vec{R}(t), t]
$$
comes from the variation of the rotation matrix.


\section{Numerical results}
We investigate the ability of the SHARC scheme to describe the effects of excitedstate dynamic Stark control. As a test system, we use a simplified model the IBr molecule. Here, a first excitation pulse $\mathrm{E}_{e}(t)$ transfers some population from the ground to an excited state. The subsequent dynamics via an avoided crossing, which is present due to $\mathrm{SOC}$, is then influenced by a control pulse $\mathrm{E}_{c}(t)$, see Fig. 1 (Transitions (a)-(c) will be explained below). The reason for the choice of this system is that for the model used, exact QD calculations are possible and we are able to compare the outcome from the different methods.

\begin{figure}
\includegraphics[width=0.47\textwidth]{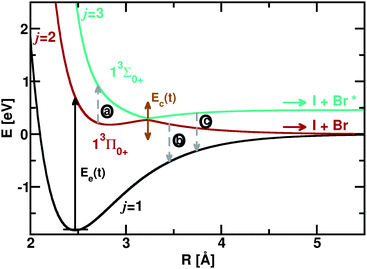}
 \caption{Potential energy curves of the IBr molecule and excitation scheme. The IBr molecule initially in the electronic ground state (black) is excited to the $1^{3} \Pi_{0+}$ excited electronic state (red) and can undergo dissociation into two different channels due to an avoided crossing introduced by SOC with the $1^{3} \Sigma_{0+}^{-}$excited state (turquoise). The control laser may shift the potential curves via NRDSE but also induce transitions at points (a)-(c), see text for explanation.}
\end{figure}

In the quantum propagations, we use the split-operator method. ${ }^{49}$ Stationary $^{2}$ states are computed via imaginary time propagation. ${ }^{50}$ Wigner distributions from the corresponding wavefunctions are employed to establish the initial conditions in the MD simulations.

For the sake of simplicity in the QD simulations, we restrict ourselves to the model potentials from Ref. 51 for IBr. They are shown in Fig. 1. As we want to model the Stark effect, we need polarizabilities and dipole moments in the direction of the laser polarization, which we assume to be the $z$-axis and therefore will omit vector arrows in the following. The static polarizabilites $\alpha_{n m}^{z z}$, permanent dipole moments $\mu_{n n}^{z}$, as well as the transition dipole moments $\mu_{n m}^{z}$ are fitted to the results published in Ref. 44 analytically. In deriving the formulas, our aim was to find a reasonable agreement with the poblished curves regardless of the number of parameters. The curves are given without SOC included, like they would be obtained from a quantum-chemical program, as:
$$
\begin{aligned}
& \alpha_{11}^{z z}=a_{1}\left(R-a_{2}\right) e^{-a_{3}\left(R-a_{2}\right)^{3}}+a_{4} \\
& \alpha_{22}^{z z}=a_{5} e^{-a_{6}\left(R-a_{7}\right)^{2}}+a_{8} e^{-a_{9}\left(R-a_{10}\right)^{2}}+a_{4} \\
& \alpha_{33}^{z z}=a_{11} e^{-a_{12}\left(R-a_{13}\right)^{2}}+a_{14} e^{-a_{15}\left(R-a_{16}\right)^{2}}+a_{4} \\
& \mu_{11}^{z}=a_{17}\left(R-a_{18}\right) e^{-a_{19}\left(R-a_{18}\right)^{2}} \\
& \mu_{22}^{z}=a_{20}\left(1-e^{-a_{21}\left(R-a_{22}\right)^{2}}\right)-a_{20}+a_{23} e^{-a_{24}\left(R-a_{25}\right)^{2}}
\end{aligned}
$$

$$
\begin{gathered}
\mu_{33}^{z}=a_{26} e^{-a_{27}\left(R-a_{28}\right)^{2}}+a_{29} e^{-a_{30}\left(R-a_{31}\right)^{2}} \\
\mu_{12}^{z}=\mu_{21}^{z}=a_{32}\left(R-a_{33}\right) e^{-a_{34}\left(R-a_{33}\right)} \\
\mu_{13}^{z}=\mu_{31}^{z}=a_{35}\left(R-a_{36}\right) e^{-a_{37}\left(R-a_{36}\right)^{2}} \\
\mu_{23}^{z}=\mu_{32}^{z}=0
\end{gathered}
$$
The parameters $a_{i}$ are listed in Table 1 . The corresponding diabatic curves are depicted in Fig. 2, left panels. They are easily adiabatized using the U $[\vec{R}(t), t]$ matrix mentioned above. If the effect of $\mathrm{SOC}$ is evaluated, the resulting curves nicely reproduce the ones given in Ref. 44 (Fig. 2, right panels).

\begin{table}
\caption{Parameters $a_{i}$ for analytical fit of $\alpha_{n}$ and $\mu_{n m}$}
\begin{tabular}{lllllllll}
\hline
Param.  & Value & Unit  & Param. & Value & Unit & Param.& Value & Unit \\\hline
a$_{1}$ & 9.28 	& \AA$^2$ 	& a$_{14}$ 	& 3.15 	& \AA$^3$ 	& a$_{27}$ 	& 5.00 	& \AA$^{-2}$ \\
a$_{2}$ 	& 2.10 	& \AA 	& a$_{1}$5 	& 3.50 	& \AA$^{-2}$ 	& a$_{28}$ 	& 2.65 	& \AA \\
a$_{3}$ 	& 0.35 	& \AA$^{-3}$ 	& a$_{16}$ 	& 3.13 	& \AA 	& a$_{29}$ 	& 0.45 	& D \\
a$_{4}$ 	& 8.20 	& \AA$^3$ 	& a$_{17}$ 	& 2.25 	& D\AA$^{-1}$ 	& a$_{30}$ 	& 5.00 	& \AA$^{-2}$ \\
a$_{5}$ 	& 105.83& \AA$^3$ 	& a$_{18}$ 	& 2.07 	& \AA 	& a$_{31}$ 	& 3.35 	& \AA \\
a$_{6}$ 	& 0.11 	& \AA$^{-2}$ 	& a$_{19}$ 	& 0.83 	& \AA$^{-2}$ 	& a$_{32}$ 	& 0.59 	& D\AA$^{-1}$ \\
a$_{7}$ 	& -2.58 & \AA 	& a$_{20}$ 	& 0.80 	& D 	& a$_{33}$ 	& 2.07 	& \AA \\
a$_{8}$ 	& 9.10 	& \AA$^3$ 	& a$_{21}$ 	& 4.00 	& \AA$^{-1}$ 	& a$_{34}$ 	& 0.67 	& \AA$^{-2}$ \\
a$_{9}$ 	& 6.43 	& \AA$^{-2}$ 	& a$_{22}$ 	& 2.25 	& \AA 	& a$_{35}$ 	& 0.38 	& D\AA$^{-1}$ \\
a$_{10}$ 	& 2.95 	& \AA 	& a$_{23}$ 	& 0.30 	& D 	& a$_{36}$ 	& 2.14 	& \AA \\
a$_{11}$ 	& 50.02 & \AA$^3$ 	& a$_{24}$ 	& 20.00 & \AA$^{-2}$ 	& a$_{37}$ 	& 3.58 	& \AA$^{-2}$ \\
a$_{12}$ 	& 5.62 	& \AA$^{-2}$ 	& a$_{25}$ 	& 2.85 	& \AA 	& 	& 	&  \\
a$_{13}$ 	& 1.68 	& \AA 	& a$_{26}$ 	& 0.90 	& D 	& 	& 	&  \\\hline
\end{tabular}
\end{table}

\begin{figure}
\includegraphics[width=0.47\textwidth]{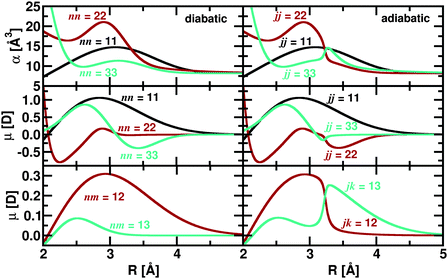}
 \caption{Polarizabilities and dipole moments of the IBr molecule in the diabatic and adiabatic representations. The curves are analytically fitted to the results from Ref. 44}
\end{figure}

The interactions between the initial diabatic states in our model system comprise SOC, dipole coupling and couplings induced by the static polarizability. To describe the interaction of the control laser with the static polarizability, the rotating wave approximation has to be employed, see, e.g., Ref. 52 . Hence, the arbitrary-couplings matrix mentioned above reads:
$$
\mathrm{H}_{n m}^{A C}[R(t)]=-\mu_{n m}[R(t)]\left(E_{\mathrm{e}}(t)+E_{\mathrm{c}}(t)\right)-\frac{1}{2} \alpha_{n}[R(t)]\left|E_{c}^{0}(t)\right|^{2}+\mathrm{H}_{n m}^{S O}[R(t)]
$$
where $E_{C}^{0}(t)$ denotes the envelope function of the control laser pulse. Depending on the field strength of the control laser, the potentials of the considered system will be shifted. Here, we investigate fields of intermediate strength (compared to weak fields, which do not alter the potentials, or to strong fields that are able to induce ionization processes and Coulomb explosion). Due to the interaction with the control laser, we create light-induced potentials (LIPs) and thus, the photodynamics is altered. In order to explore how these LIPs look like, we calculate the $V_{n}^{a}[R]$ for different field strengths. The resulting curves are plotted in Fig. 3 for field intensities of $1 \times 10^{13} \mathrm{~W} \mathrm{~cm}{ }^{-2}, 5 \times 10^{13} \mathrm{~W} \mathrm{} \mathrm{cm}^{-2}$ and the field free case for comparison.

\begin{figure}
\includegraphics[width=0.47\textwidth]{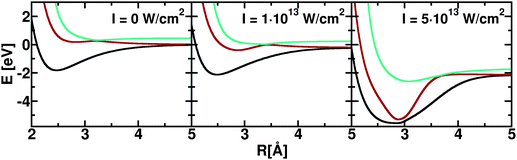}
 \caption{Changes induced in the potential energy curves of the IBr molecule by laser fields of intermediate strength. The electronic ground state (black), the $1^{3} \Pi_{0+}$ excited electronic state (red), and the $1^{3} \Sigma_{0+}^{-}$excited state (turquoise) are shifted in different manners due to their different dipole moments and polarizability.}
\end{figure}

As can be seen from the figure, the region of the avoided crossing shifts to larger interatomic distances with increasing field strength. For the present cases, the point where the two excited-state curves come closest is $R=3.23 \AA$ for $I=0 \mathrm{~W} \mathrm{~cm}{ }^{-2}, R=$ $3.40 \AA$ for $I=1 \times 10^{13} \mathrm{~W} \mathrm{~cm}{ }^{-2}$, and $R=3.68 \AA$ for $I=5 \times 10^{13} \mathrm{~W} \mathrm{~cm}{ }^{-2}$. The energy separation at these points does not change and is $0.037 \mathrm{eV}$. However, the gradient in the excited states is changed dramatically with increasing field strength. In this way, the dynamics of the system is expected to be influenced.

We now investigate how the branching ratio of the products from the different dissociation channels depicted in Fig. 1 is altered by the NRSDE. We use an excitation pulse of $493.4 \mathrm{~nm}$, a full width at half maximum of the Gaussian-shaped envelope of $50 \mathrm{fs}$ and an intensity of $1 \times 10^{13} \mathrm{~W} \mathrm{~cm}{ }^{-2}$. The control laser has a wavelength of $1.73 \mu \mathrm{m}$, similar to the experiment, ${ }^{39}$ a full width at half maximum of the Gaussian-shaped envelope of $150 \mathrm{fs}$, an intensity of $1 \times 10^{13} \mathrm{~W} \mathrm{~cm}{ }^{-2}$, and the delay $\Delta \tau$ between the pulses is scanned as indicated in Fig. 4. A negative time delay indicates the control pulse preceding the excitation pulse. In the SHARC simulations, a timestep of $0.001$ fs was used to propagate 500 trajectories unless indicated otherwise. Note that the following argumentation is also valid for lower field strengths. However, the effects will not be as pronounced as presented in the given example.

\begin{figure}
\includegraphics[width=0.35\textwidth]{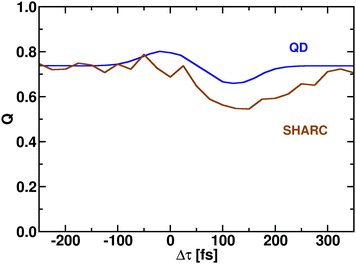}
 \caption{Branching ratio $Q$ of the different product channels as a function of delay $\Delta \tau$ between excitation and control pulse.}
\end{figure}

The branching ratio for the dissociation products from the different channels is defined as $Q=\frac{\left[I+B r^{*}\right]}{[I+B r]+\left[I+B r^{*}\right]}$. The ion yield to obtain $Q$ is represented in the QD model as the time-integrated flux ${ }^{53} F$ through the different reaction channels:
$$
F_{j}(R, t)=\int_{0}^{t}-i \frac{\psi_{j}^{*}\left(R, t^{\prime}\right) \hat{p}_{R} \psi_{j}\left(R, t^{\prime}\right)-\psi_{j}\left(R, t^{\prime}\right) \hat{p}_{R} \psi_{j}^{*}\left(R, t^{\prime}\right)}{2 M} d t^{\prime}
$$
In the SHARC calulations, the time-integrated flux is simply evaluated as the time integration of the number of trajectories at time $t^{\prime}$ at distance $R$ multiplied with the sign of the projection of the velocity vector on the direction of the flux. The timeintegrated flux is calculated at a distance $R_{F}=8 \mathrm{~A}$. In the $\mathrm{QD}$ calculations, fractions of the wavepacket at larger distances are removed employing absorbing boundary conditions as described, e.g., in Ref. 54 to avoid unphysical behavior due to limited grid size.

As can be seen from Fig. 4, results from SHARC and exact quantum dynamical simulations show the same behavior. To simulate such a curve within a surface hopping scheme is highly challenging because only a small fraction of the computed trajectories is excited from the ground state. From this small precentage, the ratio $Q$ has to be derived, i.e. the small fraction of trajectories is split once more. Therefore, a large number of trajectories has to be calculated in order to minimize statistical errors. Here, 1930 trajectories are used for each delay time. The statistical noise is small in the SHARC curve and agrees well with the results from QD. There is however an issue that we would like to point out: The calculated curves do not show the same behavior as the experimental results. ${ }^{39}$ Interestingly, curves closer to the experiment are obtained if the polarizability of the diabatic state $n=2$ is taken to be negative as done in Ref. 55. However, also in this case, the experimental curves are not reproduced correctly. Some calculations on our side indicate that the computations of Patchkovskii are correct and all static polarizabilities should be positive. for the static polarizability that can be found in textbooks ${ }^{56} \mathrm{~A}$ full understanding of the processes involved, which are much more complicated than they seem to be at first sight, lies beyond the scope of this work. Instead, we only focus on the ability of the SHARC method to provide the same results as exact quantum calculations given the same conditions and for this purpose, we used the simplified model described above. Nonetheless, in passing, we will point out some aspects that are relevant for further investigations of the NRDSE in IBr and the NRDSE in general.

In this model, the control acts in a way that by changing the potentials, the population of the excited states is either accelerated at delay times smaller than 50 fs or slowed down at delay times larger than 50 fs. This statement is exemplarily proven in Fig. 5, where the momentum is compared for delay times $\Delta \tau=0$ fs and $\Delta \tau=90$ fs to the case where no control field is present. The explanation for this acceleration and decelaration can be deduced from Fig. 6 . There, the shift in the potentials curves due to the control laser interaction is visualized for the aforemenrepresented as dot-dashed lines and the respective potentials shifted by the maximal value of the control field strength are shown as dashed lines. These two limits mark the range between which the actual LIPs are shifting during the time-dependent control pulse. An impression of how this shift occurs is given when the expectation value of the shifted potential $\left\langle V_{j}(t)\right\rangle$ at time $t$ is plotted against the expectation value of the interatomic distance $\left\langle R_{j}(t)\right\rangle$. In such a curve, a larger value of $\left\langle R_{j}(t)\right\rangle$ also signifies a larger time $t$ as we look at a dissociative process, where the bond distance is monotonically increasing. Additionally, the expectation value of the total energy $\left\langle H_{j}(t)\right\rangle$ is shown (dotted lines). The latter is not conserved during the laser interaction. The expectation values are only plotted if the population in the respective state is non-negligible.

\begin{figure}
\includegraphics[width=0.47\textwidth]{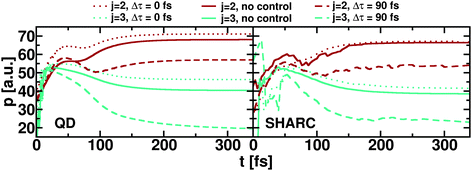}
 \caption{Mean time-dependent momentum in the excited states. Red curves correspond to $j=2$ and blue curves to $j=3$. We consider three cases: dynamics without control field (solid lines), with overlapping pulses (dotted lines) and with a time delay of $\Delta \tau=90 \mathrm{fs}$ (dashed lines). Calculations from QD and SHARC are in good agreement. A higher momentum than without control pulse is obtained for a delay of 0 fs while the momentum is lower for a delay 90 fs.}
\end{figure}

\begin{figure}
\includegraphics[width=0.47\textwidth]{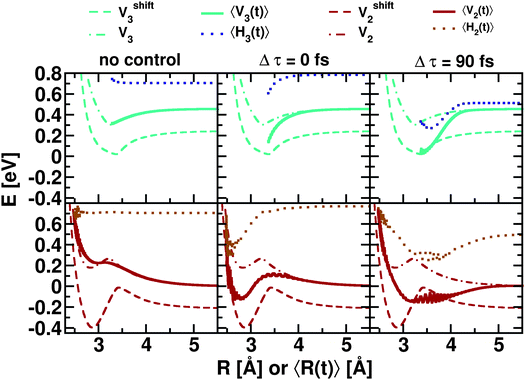}
 \caption{Effective LIPs (solid lines), unshifted potentials (dot-dashed lines), and maximally shifted LIPs (dashed lines). The latter two types serve as extrema between which an effective LIP is changing during the interaction time of the control laser. Additionally, the total energy in the respective state is plotted (dotted lines). The curves are presented for the three cases: No control field acting, delay time $\Delta \tau=0 \mathrm{fs}$, and $\Delta \tau=90 \mathrm{fs}$. The effective LIPs and the total energy are evaluated as the expectation values $\left\langle V_{j}(t)\right\rangle$ and $\left\langle H_{j}(t)\right\rangle$ at time $t$ plotted against the corresponding expectation value of the distance $\left\langle R_{j}(t)\right\rangle$ at the same time $t$. See text for further details.}
\end{figure}

Now, we regard different scenarios: When no control field is present (left panels), the effective potential mostly follows the unshifted curves. A slight deviation is visible due to the interaction with the pump laser at small $R$, i.e. at early times. The total energy is conserved. At a delay $\Delta \tau=0$ fs (middle panels), the effective potential tracks the shifted curve at small $R$ (early times) while it returns to the unshifted one at large $R$ (later times, when the control pulse is over). During the laser interaction, the kinetic energy increases faster than in the field-free case, which can be deduced from the value of the total energy. For a delay of $\Delta \tau=90$ fs (right panels), the effective potential follows the unshifted curve at small $R$ (early times), approaches the shifted curve at intermediate $R$ (intermediate times), and returns values below the shifted curve of $j=2$ due to population transfer to the state $j=3$. Due to the laser interaction, the kinetic energy is diminished in this case.

Consequently, the velocity is decreased when reaching the avoided crossing for $\Delta \tau=90 \mathrm{fs}$. According to the Landau-Zener probability, ${ }^{57}$ this case will result in diminished transitions to adiabatic state $j=3$ while for $\Delta \tau=0 \mathrm{fs}$, a higher velocity is obtained, increasing the transition probability. At higher field strengths, the kinetic energy can be decreased so much that the dissociation limit in state $j=3$ is not reached anymore (not shown).

So far, we have only examined excited states; however, interactions with the ground state may also be important. To explore the role of such interactions, we plot the probability density in the different states in Fig. 7. At around 90 fs and at a distance of $R \approx 3.8 \mathrm{~A}$, population is transfered to the ground state and starts oscillating. This can be explained as a dump process from the adiabatic state $j=3$ with the control laser corresponding to transition (c) in Fig. 1. The energy difference
between states $j=2$ and $j=3$ matches the laser frequency at (a); however, the transition dipole moment is zero between these states and a transition at this geometry is impeded. Also the transition at (b) is not possible due to a zero transition dipole moment at the corresponding bond length. Nonetheless, the control laser-although it is thought to be non-resonant in the NRDSE-may still induce resonant transitions and thus, complicate the simple picture of shifted potential curves only. Note that resonant transitions are still possible if a negative polarizability is assumed for diabatic state $n=2$. It is gratifying to see that this effect is reproduced in the SHARC simulations, as seen in Fig. 7. Also the rest of the probability distribution from the QD calculation is nicely reproduced by SHARC.

\begin{figure}
\includegraphics[width=0.47\textwidth]{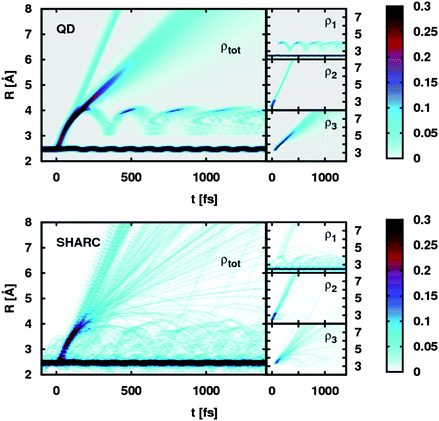}
 \caption{Probability density for a time delay of $90 \mathrm{fs}$ between excitation and control pulse. Results from SHARC are shown on the bottom, output from QD on the top. The respective panel on the left shows the total probability density and the panels on the right show the fractions of the same density belonging to the adiabatic states $j$ as indicated.}
\end{figure}

The general conclusion is that SHARC is able to describe molecular dynamics under the given, relatively strong fields. The treatment of bigger systems, which may even include molecules in solution, is straightforward by computing the potential energies "on the fly". From this perspective, SHARC is ready to be used for large systems under complex control scenarios in the presence of SOC or nonadiabatic couplings.

\section{Conclusion}

To conclude, we have shown that the new surface-hopping-in-adiabatic-representation-including-arbitrary-couplings (SHARC) algorithm is able to describe photophysical processes even at intermediate field strengths. A complex control scenario including the non-resonant dynamic Stark effect (NRDSE) was modelled by our semiclassical method, where the surface hopping probabilities are calculated in terms of a unitary transformation matrix. Thus, SHARC is able to treat all kinds of couplings in molecular systems including all degrees of freedom on the same footing.

As a control target, we have chosen the influence of the NRDSE on the branching ratio of dissociation products after photoexcitation in a model of IBr. The branching at an avoided crossing, which arises from spin-orbit coupling, gives rise to the products I + Br$^{*}$ or I + Br. The mechanism of how the branching ratio is altered was rationalized in terms of the light-induced potentials (LIPs). It was also found that resonant transitions induced by the control laser which creates the NRDSE may complicate the dynamics and are important to be considered. A comparison between SHARC and exact quantum dynamics simulations show that the respective results are in good agreement. Consequently, SHARC is able to describe laser control in complex scenarios. Therefore, it is now possible to treat photoinduced dynamics in large systems including all degrees of freedom.

\begin{acknowledgments}
This work has been supported by the Deutsche Forschungsgemeinschaft (DFG) within the project GO 1059/6-1, by the German Federal Ministry of Education and Research within the research initiative PhoNa, the Direcci{\'o}n General de Investigaci{\'o}n of Spain under Project No. CTQ2008-06760, the Friedrich-Schiller-Universit{\"a}t Jena, and a Juan de la Cierva contract, the European COST Action CM0702, and the German Academic Exchange Service (DAAD). Generous allocation of computer time at the Computer center of the Friedrich-Schiller-Universität is gratefully acknowledged.
\end{acknowledgments}

\section{References}
1 T. S. Rose, M. J. Rosker and A. H. Zewail, J. Chem. Phys., 1988, 88, 6672-6673.

2 T. S. Rose, M. J. Rosker and A. H. Zewail, J. Chem. Phys., 1989, 91, 7415-7436.

3 D. Tannor, Introduction to Quantum Mechanics: A Time-Dependent Perspective, University Science Books, Sausalito, $2006 .$

4 M. H. Beck, A. Jäckle, G. A. Worth and H. D. Meyer, Phys. Rep., 2000, 324, 1-105. 5 J. M. Bowman, T. Carrington and H. Meyer, Mol. Phys., 2008, 106, 2145-2182.

6 G. A. Worth, H. D. Meyer, H. Köppel, L. S. Cederbaum and I. Burghardt, Int. Rev. Phys. Chem., 2008, 27, 569-606.

7 A. M. Virshup, C. Punwong, T. V. Pogorelov, B. A. Lindquist, C. Ko and T. J. Martínez, J. Phys. Chem. $B, 2009,113,3280-3291$.

8 G. A. Levine, J. D. Coe, A. M. Virshup and T. J. Martínez, Chem. Phys., 2008, 347, 3-16.

9 G. A. Worth, M. A. Robb and I. Burghardt, Faraday Discuss., 2004, 127, 307-323.

10 V. A. Rassolov and S. Garashchuk, Phys. Rev. A: At., Mol., Opt. Phys., 2005, 71, 032511.

$11 \mathrm{~J}$. Li, C. Woywod, V. Vallet and C. Meier, J. Chem. Phys., 2006, 124, $184105 .$

12 R. Spezia, I. Burghardt and J. T. Hynes, Mol. Phys., 2006, 104, 903-914.

13 B. Lasorne, M. A. Robb and G. A. Worth, Phys. Chem. Chem. Phys., 2007, 9, 3210-3227.

14 D. V. Shalashilin, M. S. Child and A. Kirrander, Chem. Phys., 2008, 347, 257-262.

$15 \mathrm{~T}$. Yonehara, S. Takahashi and K. Takatsuka, J. Chem. Phys., 2009, 130,214113.

16 T. Yonehara and $\mathrm{K}$. Takatsuka, J. Chem. Phys., 2010, 132, 244102 .

$17 \mathrm{G}$. Granucci, M. Persico and A. Zoccante, J. Chem. Phys., 2010, 133, $134111 .$

18 Ab Initio Molecular Dynamics: Basic Theory and Advanced Methods, ed. D. Marx and J. Hutter, Cambridge University Press, Cambridge, 2009 .

19 N. Doltsinis and D. Marx, J. Theor. Comput. Chem., 2002, 1, 319-349.

20 J. C. Tully, J. Chem. Phys., 1990, 93, 1061-1071.

21 N. Doltsinis, in: Computational Nanoscience: Do It Yourself!, ed. J. Grotendorst , S. Blügel and D. Marx, John von Neumann Institute for Computing, NIC Series, vol. 31, Jülich, 2006 , pp. 389 409

22 B. Maiti, G. C. Schatz and G. Lendvay, J. Phys. Chem. A, 2004, 108, 8772-8781.

$23 \mathrm{~K}$. Yagi and K. Takatsuka, J. Chem. Phys., 2005, 123, 224103.

$24 \mathrm{G}$. A. Jones, A. Acocella and F. Zerbetto, J. Phys. Chem. A, 2008, 112, 9650-9656.

25 R. Mitrić, J. Petersen and V. Bonačić-Koutecký, Phys. Rev. A: At., Mol., Opt. Phys., 2009, 79, 053416 .

26 I. Tavernelli, B. F. E. Curchod and U. Rothlisberger, Phys. Rev. A: At., Mol., Opt. Phys., $2010, \mathbf{8 1}, 052508 .$

27 M. Richter, P. Marquetand, J. González-Vázquez, I. Sola and L. González, J. Chem. Theory Comput., 2011, 7, 1253-1258.

28 P. Brumer and M. Shapiro, Annu. Rev. Phys. Chem., 1992, 43, 257-282.

29 R. J. Gordon and S. A. Rice, Annu. Rev. Phys. Chem., 1997, 48, 601-641.

30 S. A. Rice, Adv. Chem. Phys., 1997, 101, 213-283.

31 D. J. Tannor, R. Kosloff and A. Bartana, Faraday Discuss., 1999, 113, 365-383.

32 S. A. Rice and M. Zhao, Optical Control of Molecular Dynamics, Wiley, New York, 2000 .

33 T. Brixner, N. H. Damrauer and G. Gerber, Adv. At., Mol., Opt. Phys. 2001, 46, 1-54.

34 M. Shapiro and P. Brumer, Rep. Prog. Phys., 2003, 66, 859-942.

35 M. Shapiro and P. Brumer, Principles of Quantum Control of Molecular Processes, Wiley, New York, $2003 .$

36 C. Daniel, J. Full, L. González, C. Lupulescu, J. Manz, A. Merli, Stefan Vajda and L. Wöste, Science, 2003, 299, 536-539.

37 P. Nuernberger, G. Vogt, T. Brixner and G. Gerber, Phys. Chem. Chem. Phys., 2007, 9 , 2470-2497.

$38 \mathrm{~V}$. Engel, C. Meier and D. J. Tannor, Adv. Chem. Phys., 2009, 141, 29-101.

39 B. J. Sussman, D. Townsend, M. Y. Ivanov and A. Stolow, Science, 2006, 314, 278-281.

$40 \mathrm{~J}$. González-Vázquez, I. R. Sola, J. Santamaria and V. S. Malinovsky, Chem. Phys. Lett., $2006, \mathbf{4 3 1}, 231-235$.

41 B. Y. Chang, S. Shin, J. Santamaria and I. R. Sola, J. Chem. Phys., 2009, 130, $124320 .$

42 B. Y. Chang, S. Shin and I. R. Sola, J. Chem. Phys., 2009, 131, $204314 .$

43 M. Wollenhaupt, A. Präkelt, C. Sarpe-Tudoran, D. Liese and T. Baumert, J. Opt. B: Quantum Semiclassical Opt., 2005, 7, S270–S276.

44 S. Patchkovskii, Phys. Chem. Chem. Phys., 2006, 8, 926-940.

45 L. Verlet, Phys. Rev., 1967, 159, 98-103.

46 L. Verlet, Phys. Rev., 1968, 165, 201-214.

47 M. Barbatti, A. J. A. Aquino, J. J. Szymczak, D. Nachtigallová, P. Hobza and H. Lischka, Proc. Natl. Acad. Sci. U. S. A., 2010, 107, 21453-21458.

$48 \mathrm{~J}$. González-Vázquez and L. González, ChemPhysChem, 2010, 11, 3617-3624

49 M. D. Feit, J. A. FleckJr. and A. Steiger, J. Comput. Phys. 1982, 47, 412-433.

$50 \mathrm{R}$. Kosloff and H. Tal-Ezer, Chem. Phys. Lett., 1986, 127, 223-230.

$51 \mathrm{H}$. Guo, J. Chem. Phys. 1993, 99, 1685-1692.

52 T. Seideman, J. Chem. Phys., 1997, 106, 2881.

53 P. Marquetand, S. Gräfe, D. Scheidel and V. Engel, J. Chem. Phys., 2006, 124, 054325/1-7.

54 P. Marquetand and V. Engel, Chem. Phys. Lett., 2006, 426, 263-267. 55 B. J. Sussman, M. Y. Ivanov and A. Stolow, Phys. Rev. A: At., Mol., Opt. Phys., 2005, 71, 051401 .

56 C. Cohen-Tannoudji, B. Diu and R. Laloë, Quantum Mechanics, Vol. 2, Wiley, New York, $1977 .$

57 P. Marquetand and V. Engel, Chem. Phys. Lett., 2005, 407, 471-476.

58 M. Thachuk, M. Y. Ivanov and D. M. Wardlaw, J. Chem. Phys., 1996, $\mathbf{1 0 5}, 40944104 .$

59 N. Shenvi, J. Chem. Phys., 2009, 130, $124117 .$

\end{document}